\documentclass[aps,prd,twocolumn,superscriptaddress]{revtex4-1}

\usepackage{natbib}
\usepackage{setspace}
\usepackage{textgreek}
\usepackage[dvipsnames]{xcolor}
\usepackage{amsmath}
\usepackage{enumitem}
\usepackage{ragged2e}
\usepackage{float}
\usepackage{graphicx}
\usepackage{xcolor}
\usepackage{dcolumn}
\usepackage{bm}
\usepackage[utf8]{inputenc}
\usepackage{bbm}

\usepackage{changes}
    \colorlet{Changes@Color}{blue}

\usepackage{hyperref}
\hypersetup{%
  colorlinks=true,
  citecolor=blue,
  urlcolor=blue,
  linkcolor=blue,
  }

\begin{document}

\title{Nonlinear spectroscopy of excitonic states in transition metal dichalcogenides}

\author{Yaroslav V. Zhumagulov}
\thanks{These authors contributed equally to this work.}
\affiliation{University of Regensburg, Regensburg 93040, Germany}
\affiliation{ITMO University, St. Petersburg 197101, Russian Federation}

\author{Vyacheslav D. Neverov}
\thanks{These two authors contributed equally}
\affiliation{National Research Nuclear University MEPhI, Moscow 115409, Russian Federation}
\affiliation{ITMO University, St. Petersburg 197101, Russian Federation}

\author{Alexander E. Lukyanov}
\affiliation{National Research Nuclear University MEPhI, Moscow 115409, Russian Federation}
\affiliation{ITMO University, St. Petersburg 197101, Russian Federation}

\author{Dmitry R. Gulevich}
\affiliation{ITMO University, St. Petersburg 197101, Russian Federation}

\author{Andrey V. Krasavin}
\affiliation{National Research Nuclear University MEPhI, Moscow 115409, Russian Federation}

\author{Alexei Vagov}
\affiliation{ITMO University, St. Petersburg 197101, Russian Federation}
\affiliation{Theoretische Physik III, Universität Bayreuth, 95440 Bayreuth, Germany}
\affiliation{National Research University Higher School of Economics, 101000 Moscow, Russia}

\author{Vasili Perebeinos}
\email{vasilipe@buffalo.edu}
\affiliation{Department of Electrical Engineering, University at Buffalo, The State University of New York, Buffalo, New York 14260, USA}

\date{\today}

\begin{abstract}
 Second-harmonic generation (SHG) is a well-known nonlinear spectroscopy method to probe electronic structure, specifically, in transition metal dichalcogenide (TMDC) monolayers. This work  investigates the nonlinear dynamics of a strongly excited TMDC monolayer by solving the time evolution equations for the density matrix. It is shown that the presence of excitons qualitatively changes the nonlinear dynamics leading, in particular, to a huge enhancement of the nonlinear signal as a function of the dielectric environment. It is also shown that the SHG polarization angular diagram and its dependence on the driving strength are very sensitive to the type of  exciton state. This sensitivity suggests that SHG spectroscopy is a convenient tool for analyzing the fine structure of excitonic states.
\end{abstract}

\maketitle

\section{Introduction}

Second-harmonic generation  (SHG)~\cite{Boyd2020} is a powerful tool for studying optical properties of a variety of materials, including semiconductors~\cite{Ghimire2010, Chin2001, Belyanin2005}, molecules~\cite{Prasad1991}, carbon nanotubes~\cite{DeDominicis2004, Murakami2009, Kono2013}, and layered transition metal dichalcogenide (TMDC) structures~\cite{Kumar2013,Malard2013,Yin2014,Clark2014,Hsu2014,Janisch2014,Jiang2014,Liu2016,Syntjoki2017,Autere2018,Mennel2018,Mennel2019,Stiehm2019,MiltosMaragkakis2019, Lin2019,Zhang2020,Khan2020,Ho2020}. The nonlinear nature of SHG allows one to probe material characteristics on a level that usually evades linear spectroscopy methods. For example, the sensitivity of SHG measurement to spatial and time-reversal symmetries makes it handy for uncovering phenomena that are out of the reach of more traditional optical methods~\cite{HEINZ1991}, including  magnetic ordering~\cite{Fiebig1998, Shree2020} and hidden phase transitions~\cite{Fiebig2005}. Recently SHG spectroscopy has been applied to map strain profiles with the spatial resolution surpassing the optical diffraction limit~\cite{Mennel2018, Mennel2019}.

In TMDC monolayers~\cite{Ma2020}, the SHG method has been  applied to investigate the symmetry of crystal structures~\cite{Kumar2013, Malard2013, Li2013,Yin2014,Clark2014, Hsu2014,MiltosMaragkakis2019},  detect charged molecules~\cite{Yu2016},  map strains~\cite{Mennel2018, Mennel2019, Khan2020}, and  probe valley polarization~\cite{Wehling2015, Hipolito2017, Ho2020}. SHG signal is very sensitive to electronic excitations, making it a powerful tool for studying  the band structure and interband transitions. The SHG is a unique tool to fill the gap left by the Raman and photoluminescence spectroscopy, and it is well suited to study atomic and electronic structures of two-dimensional (2D) layered TMDC systems~\cite{Zhang2020}. Experiments with layered $\text{WSe}_2$ on hexagonal boron nitride (h-BN) substrates~\cite{Lin2019}, $\text{WS}_2$~\cite{Janisch2014}, $\text{MoS}_2$ monolayers~\cite{Mennel2018} and bilayers~\cite{Jiang2014}, and a 2D GaSe crystal~\cite{Zhou2015} revealed a conventional six-leaf pattern of the SHG signal angular dependence, which is commonly employed to determine the orientation of the monolayer crystals~\cite{Kumar2013, Malard2013, Li2013,Yin2014,Clark2014, Hsu2014,MiltosMaragkakis2019}. Distortions of that symmetry, e.g., by an applied tensile strain, give rise to a distorted SHG angular dependence~\cite{Mennel2019}.

We show here that the SHG signal's sensitivity to the excitation field intensity, frequency, and polarization can be used to probe the nature of electronic states in TMDC materials on a  much more detailed level than linear spectroscopy allows. We find that unlike most materials, in TMDC, both linear and quadratic terms in the vector potential of the excitation laser light must be included to describe correctly nonlinear light-matter interaction and nonlinear dynamics. In particular, we find that the competition between the linear and quadratic terms as a function of the laser power leads to peculiar changes in the polarization diagrams of the SHG pattern. Unlike the linear response spectra, the interpretation of the SHG signals involves theoretical analysis that cannot be limited to calculating the transition energies and rates but requires one to investigate the nonlinear dynamics of the system, which is a much more complex problem. In some cases, the analysis can be simplified by employing the dynamical perturbation theory to calculate the second harmonic. However, the perturbative approach fails in the most physically interesting case of strong excitations and highly nonlinear dynamics or when the higher-order harmonics are essential. It is also not convenient when one has a mixture of excitations of different nature. In those cases, the complete nonlinear dynamical problem must be solved. For TMDC monolayers, it is often solved for single-particle excitations treated within the semi-classical approximation.  This approach is well justified, e.g., when the driving field of extra-strong intensity produces many high harmonics.

In many other relevant situations, such as SHG, the field strength is not strong enough to ionize excitons, and single-particle approximation may become inadequate, although the dynamics is still nonlinear. In 2D materials, this regime is easily accessible because of the strong Coulomb interaction enhancing the many-body effects. It facilitates the formation of tightly bound many-particle excitonic complexes, manifested in the linear optical spectra of 2D TMDC structures~\cite{Heinz2010,Wang2010,Chernikov2014, Seyler2015, Wang2015, Wang2018}. The exciton-related effects should also be visible in nonlinear dynamics, in particular, SHG. However, investigations of the nonlinear dynamics associated with the excitonic states are currently in an infant stage.  Contemporary research focuses mainly on the linear response~\cite{Cheiwchanchamnangij2012, Ramasubramaniam2012,Louie2013,Berkelbach2013,Louie2015,Zhumagulov2020_1,Zhumagulov2020,zhumagulov2021electrostatic}. Analysis of the nonlinear effects such as SHG  mainly concentrates on perturbative calculations~\cite{Trolle2014,Grning2014,Glazov2017,Kolos2021}. A more elaborate investigation of the exciton dynamics in TMDC can be done by solving  dynamics equations for the pertinent elements of the density matrix~\cite{Richter2010} obtained using the dynamics control truncation (DCT) approximation~\cite{Axt2004}.

In this work, we investigate the role of exciton states in the SHG of TMDC monolayers. Using the solution for the Liouville - von Neumann (LvN) equation, we obtain SHG angular polarization diagrams and study their dependence on the frequency and intensity of the excitation pulse. The results reveal an extraordinary sensitivity of the SHG signal to the type of exciton states. Our findings are general, and similar SHG polarization diagrams are obtained for monolayers of common TMDC's  ($\text{MoS}_2$, $\text{MoSe}_2$, $\text{WS}_2$, and $\text{WSe}_2$)  with a qualitatively similar two valley band structure.  Our results offer the tantalizing possibility of SHG spectroscopy of exciton states that can be used to probe and detect finer details of excitonic states that conventional optical methods cannot capture.

\begin{figure*}
    \centering
    \includegraphics[width=1.0\textwidth]{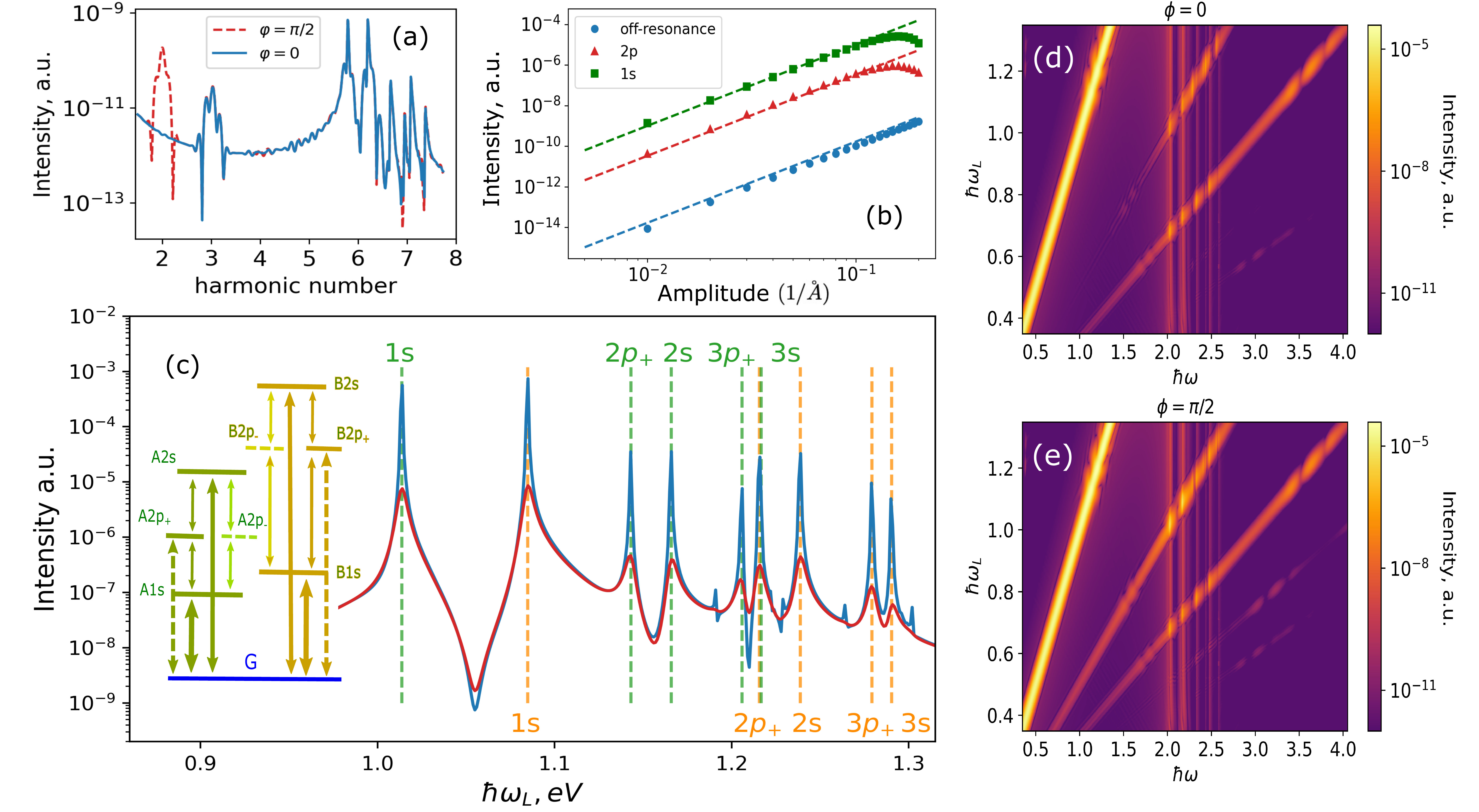}
    \caption{(a) Spectral intensity of the nonlinear response $I_{\omega_L,\omega}$ calculated at  $f = 0.5\AA^{-1}$ and $\hbar \omega_L = 0.35$eV (off-resonance) for polarization angle $\phi = 0$ (blue) and $\phi = \pi/2$ (red). (b)  SHG intensity  $I_{\omega_L,2 \omega_L}$  calculated for the off-resonant driving with $\hbar \omega_L = 0.35$eV  (blue circles), in resonance with the $A1s$ exciton state (green squares), and in resonance with the $A2p_+$ exciton state (red triangles).  Dashed lines show the  perturbation theory result $I \propto f^4$. (c) SHG intensity as a function of $\hbar \omega_L$ calculated at $f=0.1\AA^{-1}$ without dephasing ($\gamma =0$, blue line) and with dephasing ($\gamma = 10$meV, red line). A and B excitons are marked by green and orange dashed lines. The inset is a schematic  energy diagram (not to scale) of optically active low-energy excitons in a $\text{MoS}_{2}$ monolayer. Solid arrows mark transitions allowed in a rotationally invariant system; dashed lines are warping mediated transitions. (d-e) Color density high harmonic intensity spectrum as a function of the driving frequency $I_{\omega_L, \omega}$ calculated for two polarization angles  $\phi = \pi/2$ and $0$, respectively.
    }
    \label{fig:intensity}
\end{figure*}

\section{Dynamics of exciton states}

The analysis of the harmonic generation is done by employing  the formalism of the density matrix where we solve the LvN equation,
\begin{equation}
\frac{d \varrho}{d t}  = - \frac{\mathbbm{i} }{\hbar} \big[ H, \varrho \big] +  {\cal L} [\varrho],
\label{eq:LvN}
\end{equation}
where $\varrho$ is the density matrix, $H$ is the Hamiltonian of the system, and the  non-Hamiltonian contribution ${\cal L}$ accounts for the losses. Since we are interested in the exciton contribution to the dynamics, we consider the two-particle Hamiltonian. It contains two contributions $H = H_{ex}  + H_{f}$ where $H_{ex}$ is the part describing the exciton states; and $H_{f}$ is the interaction between the excitons and the driving field. The losses are taken into account by using the phenomenological Lindblad approach where the non-Hamiltonian part of the LvN equation reads as
\begin{equation}
{\cal L}=\sum_{a} \gamma_a \left[ {\cal L}_a \rho {\cal L}_a^\dagger-\frac{1}{2}\left(\rho {\cal L}_a^\dagger {\cal L}_a + {\cal L}_a^\dagger {\cal L}_a \rho\right)\right],
\end{equation}
with ${\cal L}_a$ and $\gamma_a$ being an operator and rate corresponding to a loss channel $a$. In this work, we account only for the pure dephasing mechanism, for which the operators ${\cal L}_a=|a\rangle\langle a|$  are diagonal in the basis of exciton states $a$.  
The resulting LvN equation is solved using a basis of two-particle exciton eigenstates, which are in turn calculated using conduction and valence band single-particle states of a TMDC material as a basis.

\subsection{Single-particle states}

Single-particle states are obtained from the massive Dirac model Hamiltonian, which takes into account the trigonal warping and spin-orbit coupling  \cite{Xiao2012,Kormnyos2013,Kormnyos2015}. It yields a good approximation for the low-lying states of a TMDC monolayer.  The  Hamiltonian reads as
\begin{align}
H_{0}({\bf k})  & = \frac{\Delta}{2} \sigma_z  +  \hbar v_F \tau (  k_x \sigma_x +   \tau k_y \sigma_y)   \notag \\
& + \frac{ \tau s}{2}  \big\{  \lambda_c  (\sigma_z + \sigma_0) -  \lambda_v ( \sigma_z - \sigma_0) \big\}  \notag \\ & + \kappa \left\{  \left(k_x^2 - k_y^2 \right) \sigma_x + 2 \tau k_x k_y \sigma_y \right\} ,
\label{eq:Dirac}
\end{align}
where  $\sigma_i$ are the Pauli matrices ($\sigma_0$ is the unity matrix),  ${\bf k} = (k_{x},k_y)$ are the electron momentum components, $\tau = \pm 1$ is the valley index, $s=\pm 1$ denotes charge carrier spin,  $v_F$ is the Fermi velocity,  $\Delta$ is the gap between the conduction and the valence bands, and  $\lambda_{c,v}$ are the spin splittings of the conduction $c$ and valence $v$ bands.  The Fermi velocity is given as $\hbar v_F=\sqrt{\Delta/2m}$ \cite{zhumagulov2021electrostatic}, and the trigonal warping constant is determined as $\kappa=-\sqrt{3}/24 \hbar v_F a$ \cite{Kormnyos2013,Taghizadeh2019}. Finally, we assume the following model to describe the dependence of the bandgap on the dielectric environment \cite{Cho2018}
\begin{align}
    \Delta & =\Delta_{0}+\frac{e^2}{2\epsilon d}\Big[\frac{L_{2}+L_{1}}{\sqrt{L_{2}L_{1}}}\text{tanh}^{-1}(\sqrt{L_{2}L_{1}}) \notag \\ & -\text{ln}(1-L_{2}L_{1})\Big], \quad L_{i}=\frac{\epsilon-\epsilon_{i}}{\epsilon+\epsilon_{i}},
\end{align}
where $\epsilon$ is the bulk dielectric constant of the TMDC material,  $\Delta_0$ is its bulk bandgap, $\epsilon_{i=1,2}$ are dielectric constants of the lower and upper dielectric environment, and $d$ is the monolayer thickness. Eigenstates of the Dirac Hamiltonian are classified as quasi-electrons $|c {\bf k}\rangle$ with energies $\varepsilon_{c{\bf k}}$ above the gap and quasi-holes  $|v{\bf k}\rangle$  with energies $\varepsilon_{v{\bf k}}$  below the gap. Indices $c$ and $v$ denote all state quantum numbers except for the quasimomentum~${\bf k}$. Monolayers of semiconductors $\text{MoS}_2$, $\text{MoSe}_2$, $\text{WS}_2$, and $\text{WSe}_2$ have a similar crystal configuration and qualitatively similar energy dispersion of the lowest energy single-particle states.   The effective Dirac model parameters for those materials are obtained by fitting results of the first-principle band structure calculations and are summarized in Table~\ref{tab:parameters}.

\begin{table}[]
    \centering
    \begin{tabular}{|c|c|c|c|c|c|c|c|}
        \hline
           & $a$  & $d$   & $\epsilon$ & $\Delta_0$ & $m$  & $\lambda_c$ & $\lambda_v$ \\
         \hline
         \hline
         MoS$_2$&  3.185 & 6.12 & 16.3 & 2.087 & 0.520 & -1.41 &  74.60\\
         \hline
         MoSe$_2$& 3.319 & 6.54 & 17.9 & 1.817 & 0.608 & -10.45 &  93.25\\
          \hline
         WS$_2$ &   3.180  & 6.14 & 14.6 & 2.250 & 0.351 & 15.72  &  213.46\\
          \hline
         WSe$_2$ &  3.319 & 6.52 & 16.0 & 1.979 & 0.379 & 19.85  &  233.07\\
         \hline
    \end{tabular}
    \caption{Model parameters for TMDC monolayers. Lattice constants $a$ (\AA), effective masses $m$, and spin-orbit coupling constants $\lambda_{c,v}$ (meV) are taken from Ref. \cite{Zollner2019}, monolayer thicknesses $d$ (\AA) and static dielectric constants $\epsilon$ are from Ref. \cite{Laturia2018}, and $\Delta_0$ (eV) is from Ref. \cite{Zhang2016}. }
    \label{tab:parameters}
\end{table}

With additional contributions due to the trigonal warping, the minimal coupling model describes the interaction between quasiparticles and the driving electromagnetic field.  We use a common assumption that the external field has a very large wavelength compared  with other system characteristic sizes. Using this assumption, we obtain the following interaction Hamiltonian for the states of momentum  ${\bf k}$~\cite{Sipe1993}:
\begin{equation}
\label{eq:H_f}
H_{f} ({\bf k})  =
-\frac{e}{c \hbar}\sum_{\alpha}p^{\alpha}_{\bf k} A_{\alpha} + \frac{e^2}{2c^2 \hbar^2} \sum_{\alpha\beta} q^{\alpha\beta}_{\bf k}  A_{\alpha}A_{\beta},
\end{equation}
where $\textbf{A}$ is the field  vector potential  $e$ is the electron charge, $c$ is the speed of light, and the coefficients of the linear and quadratic interaction terms are obtained as derivatives:
\begin{equation}
\label{eq:p_q}
p^{\alpha}_{\bf k}  = \frac{\partial H_{0} ({\bf k})  }{\partial k_{\alpha}}, \quad q^{\alpha\beta}_{\bf k} =  \frac{\partial^2 H_{0}({\bf k})   }{\partial k_{\alpha} \partial  k_{\beta}}.
\end{equation}
 Notice that unlike the Schrodinger equation with the quadratic dispersion, the term with the second power of the field facilitates transitions between single-particle states and thus cannot be neglected.

\subsection{Exciton states}

The full many-body Hamiltonian with the Coulomb interaction between electrons and holes is projected onto a basis of two-particle states $| c v \rangle = c_{c{\bf k}}^\dagger  d_{v{\bf k}}^\dagger | 0 \rangle$.  Using the basis of these states, excitons are obtained by solving the Bethe -- Salpeter equation (BSE)~\cite{Rohlfing2000} that takes into account screening due to the environment that embeds the monolayer \cite{Zhumagulov2020,Zhumagulov2020_1}. Solving the BSE is equivalent to finding eigenstates of the effective two-particle Hamiltonian $H_{ex}$ defined by its matrix elements as
\begin{align}
\label{eq:exciton_Hamiltonian}
 \langle c^\prime v^\prime |  H_{ex}  | c v \rangle   = (\varepsilon_{c} -\varepsilon_{v}) \delta_{c}^{c^\prime}   \delta_{v}^{ v^\prime} - W_{v^\prime c}^{vc^\prime } + V_{v^\prime c}^{c^\prime v},
\end{align}
where $\varepsilon_{ck,vk}$ are single-particle energies, and  $W$ and $V$ are the screened and bare Coulomb potentials. The latter is defined as $ V^{ab}_{cd} = V(\textbf{k}_a -\textbf{k}_c) \langle u_c | u_a \rangle \langle u_d | u_b \rangle$, with $\langle u_c| u_a \rangle$ being the overlap of the single-particle Bloch states, and $V ({\bf q}) = 2\pi e^2/ q$. In Eq.~(\ref{eq:exciton_Hamiltonian}), momentum index $\textbf{k}$ is absorbed in indices $c$ and $v$ for brevity. In the screened potential one changes $V({\bf q})$ for
\begin{align}
\label{eq:Coulomb_screened}
   W({\bf q})=
    \frac{2\pi e^2}{ q} \begin{cases}
     \epsilon_{env} ^{-1} (1+r_0 q)^{-1},  \, q \in \text{intra-valley}; \\
   \epsilon^{-1}, \quad   q \in \text{inter-valley}, \\
    \end{cases}
\end{align}
so that the intravalley screening (small $q$) is described by the Rytova-Keldysh potential~\cite{rytova1967the8248,keldysh}  whereas the intervalley screening is reduced to the bulk dielectric constant $\epsilon$ ~\cite{zhumagulov2021electrostatic}. Here $  \epsilon_{env}=(\epsilon_1+\epsilon_2)/2$ is determined by the dielectric environment, and  the screening length is $r_0=\epsilon d/2$. We present results for $\epsilon_{env}=1$ throughout this paper, unless otherwise stated.

The large wavelength assumption for the driving field implies that the excitation does not change the total momentum, and thus only zero-momentum  particle-hole pairs contribute to exciton states
\begin{align}
| \Psi_a \rangle =\sum_{{\bf k} }  \sum_{cv}  X_{{\bf k} cv}^a c_{{\bf k} c}^\dagger  d_{{\bf k} v}^\dagger | 0 \rangle,
\label{eq:exciton_eigenstates}
\end{align}
where  $X_{{\bf k} cv}^a$ are eigenvectors of the Hamiltonian~(\ref{eq:exciton_Hamiltonian}). In the numerical calculations, we use a mesh of ${480\times480\times1}$ in the Brillouin zone. The momentum cut-off $k_c$ is introduced to restrict the number of single-particle states near the $K$ and $K'$ valleys. The wavevector cut-off $k_c$ determines the energy cut-off of the single-particle states contributing to the basis-set of the Bethe-Salpeter equation for excitons. A chosen value of $k_c=0.4 \AA^{-1}$ is  sufficient for numerical convergence of the exciton energies and dipole transitions.
All phases must be treated consistently~\cite{Sipe2000,Rohlfing2000} in the exciton wavefunctions in Eq.~(\ref{eq:exciton_eigenstates}) to correctly describe the interference effects in the matrix elements entering the LvN equation.

The interaction Hamiltonian in the exciton states representation is found as
\begin{align}
\label{eq:Hf_exc}
H_{f} = - \frac{e}{c \hbar} \sum_{\alpha} P^{\alpha} A_{\alpha}  + \frac{e^2}{2 c^2 \hbar^2}  \sum_{\alpha\beta}
 Q^{\alpha\beta} A_{\alpha} A_{\beta},
\end{align}
where operators  $P_\alpha $ and $Q_{\alpha \beta}$ are obtained by calculating field-induced matrix elements of the  interaction Hamiltonian~(\ref{eq:H_f})  for transitions between exciton eigenstates in Eq.~(\ref{eq:exciton_eigenstates}). For  transitions that involve the ground state without an exciton (index $b=0$) one obtains the matrix elements as
\begin{align}
& P^{\alpha}_{a0} =\frac{1}{N} \sum_{{\bf k} }  \sum_{cv}   p^{\alpha}_{{\bf k} cv} X_{{\bf k} cv}^{a*}, \notag \\
&Q^{\alpha\beta}_{a0}   =\frac{1}{N} \sum_{{\bf k} }  \sum_{cv} q^{\alpha\beta}_{{\bf k} cv} X_{{\bf k} cv}^{a*},
\end{align}
where $N$ is the number of the ${\bf k} $-mesh points,  while for transitions between different exciton states (of the same total momentum), one gets
\begin{align*}
& P^{\alpha}_{ab}  =\sum_{{\bf k} }   \sum_{c c^\prime v v^\prime} X_{{\bf k} cv}^{a*} (p^{\alpha}_{{\bf k}  cc^\prime} \delta_{v}^{v^\prime} - p^{\alpha}_{{\bf k} vv^\prime} \delta_{c}^{c^\prime} ) X_{{\bf k} c^\prime v^\prime }^{b}, \notag \\
&Q^{\alpha\beta}_{ab}  =\sum_{{\bf k} }   \sum_{cc^\prime v v^\prime} X_{{\bf k} cv}^{a*} (q^{\alpha\beta}_{{\bf k}  c c^\prime}\delta_{v}^{v^\prime}  - q^{\alpha\beta}_{ {\bf k}  vv^\prime} \delta_c^{c^\prime} ) X_{{\bf k}  c^\prime v^\prime }^{b }.
\end{align*}
 The single-particle transition matrix elements in these expressions are  given as
\begin{align}
 p^{\alpha}_{{\bf k} r r^\prime }  = \langle r |  p^{\alpha}_{\bf k}  | r^\prime  \rangle, \quad  q^{\alpha \beta }_{{\bf k}  r r^\prime }  = \langle r |  q^{\alpha \beta }_{\bf k}  | r^\prime \rangle,
\end{align}
where $r$ denotes $c$ or $v$ states. In the numerical calculations, we use 48 lowest energy excitonic states (twelve four-fold degenerate states), which is sufficient for the numerical convergence of the results.

\begin{figure*}
    \centering
    \includegraphics[width=0.99\textwidth]{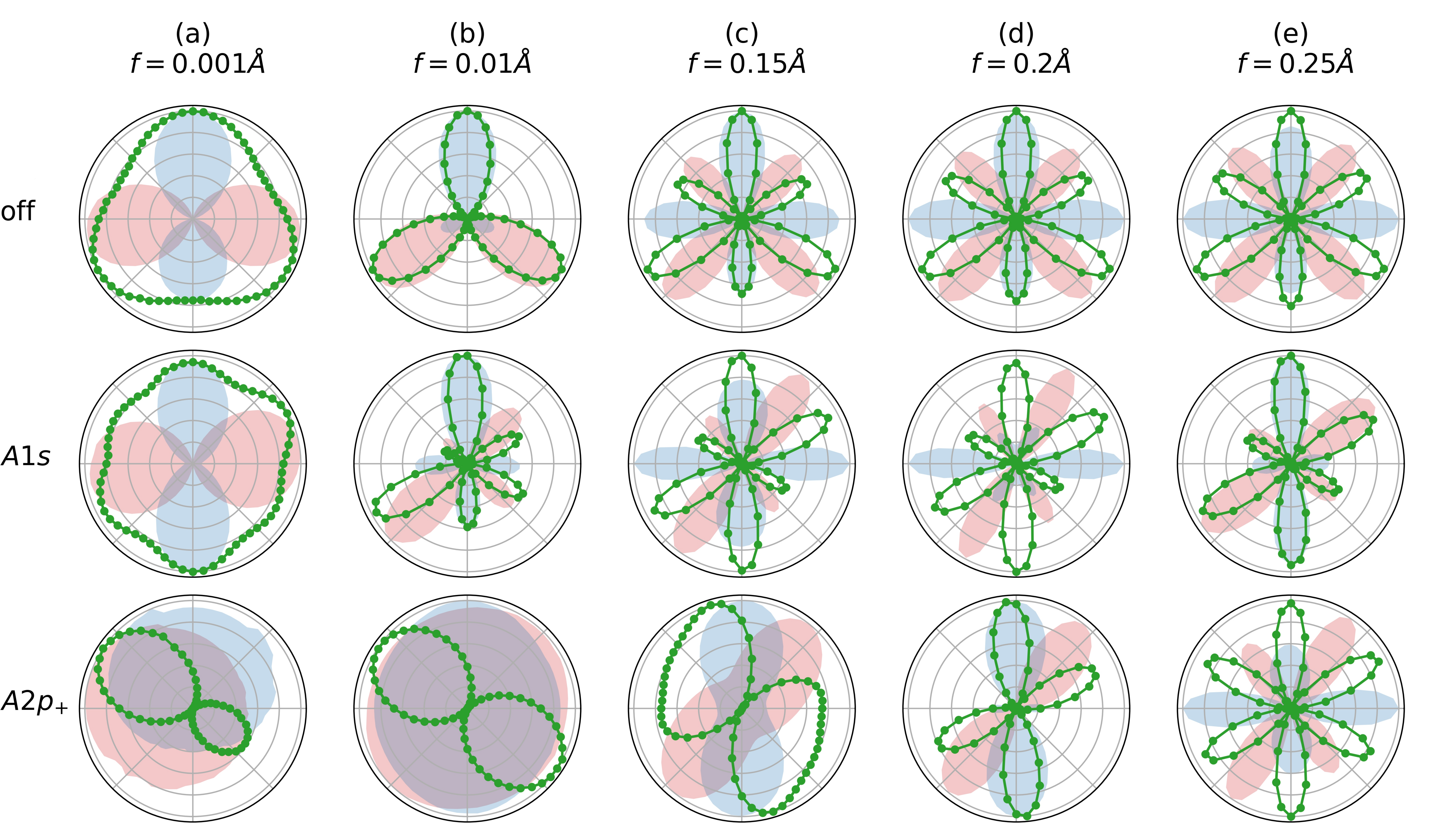}
    \caption{Polarization angle dependence of SHG intensity (green points) and polarization components $|\bar P^x|^2$ (red shading) and $|\bar P^y|^2$ (blue shading) calculated at $f = 10^{-3}\rm\;\AA^{-1}$ (a), $f=10^{-2}\rm\;\AA^{-1}$ (b), $f=0.15\rm\;\AA^{-1}$ (c), $f=0.2\rm\;\AA^{-1}$ (d), and $f=0.25\rm\;\AA^{-1}$ (e).  Top row:  off-resonant excitation with $\hbar \omega_L = 0.35\;$eV; middle row:    $\hbar \omega_L = 1.015\;$eV  is in resonance with the energy of the $A1s$ exciton; bottom row:  $\hbar \omega_L = 1.14\;$eV is in resonance with the $A2p_+$ exciton. The results are normalized to their maximal values.
    }
    \label{fig:SHG_polarization}
\end{figure*}

\subsection{Nonlinear dynamics and harmonics generation}

The dynamics of the system is obtained by solving the LvN equation~(\ref{eq:LvN}) in the presence of the driving pulse. The vector potential of the driving field is assumed to be tangential to the TMDC monolayer with the spatial components ${\bf A}(t) = A(t) (\cos(\phi), \sin (\phi), 0)$, where angle $\phi$ is measured from the zig-zag axis of the monolayer TMDC. The pulse is  monochromatic with the frequency $\omega_L$ and the Gaussian envelope function
\begin{align}
A(t) = \frac{c \hbar f}{e} \cos(\omega_L t) e^{-(t-t_0)^2/{2 \sigma^2}},
\end{align}
with the following parameters \cite{Liu2016}: $t_0 = 3 \sigma$ and pulse duration $6\sigma$, where  $\sigma = 25\rm\;fs$. The electric field amplitude is found from the vector potential in the usual way. When the envelope function varies slowly in comparison with the oscillation period, the field strength is approximately ${\bf E}  \approx  {\bf A} \omega_L /c$, which is related to the driving amplitude $f$ as $E = \hbar \omega_L f/e$ and the laser intensity $I=c E^2/ 8 \pi$ (in CGS units). Using these quantities one can calculate the laser field intensity in the conventional units as $I=13.3 (\hbar \omega_L f)^2$~TW/cm$^2$, where the laser  energy $\hbar \omega_L$ and the amplitude $f$ are given in the units of eV and \AA$^{-1}$, correspondingly.  

As described above, the losses in the system are modeled by the pure dephasing mechanism. For simplicity, all excitonic states are assumed to have the same  dephasing rate $\gamma_a = \gamma = 10$ meV. We note that our conclusions do not depend qualitatively on the details of the loss mechanism.

The LvN Eq. (\ref{eq:LvN}) is solved using Quantum Toolbox in Python (QuTiP) \cite{Johansson2012, Johansson2013}.
The obtained solution for the density matrix is then used to calculate the time evolution of the polarization operator by taking the trace
\begin{equation}
\bar P^\alpha (t)  = {\rm Tr}\left[ \varrho P^\alpha \right].
\end{equation}
The second and higher harmonics are extracted separately from the Fourier components of $\bar P^{\alpha}(t)$. In the calculations, we use the maximal propagation time of  $t_{max} = 2.5$ ps to ensure that the steady-state is achieved. Polarization components are susceptible to the polarization angle $\phi$ of the excitation pulse. When the emitted light is detected at the same polarization angle, which is a  typical experimental setup~\cite{Kumar2013, Malard2013, Li2013,Liu2016,Yin2014,MiltosMaragkakis2019}, the spectral intensity of the detected signal is proportional to
\begin{equation}
    I_{\omega_L,   \omega }  \propto {|\bar P^x_{\omega_L, \omega  } \cos(\phi) + \bar P^y_{\omega_L, \omega  } \sin(\phi) |}^2.
    \label{eq:intensity_polarization}
\end{equation}
The intensities of second and higher harmonics are given by $ I_n= I_{\omega_L,   n \omega_L}$, where $n$ is an integer.

\section{Numerical results}

\subsection{Exciton-mediated nonlinear spectrum}

A typical nonlinear spectrum of a standalone $\text{MoS}_2$ monolayer in Fig.~\ref{fig:intensity}a is calculated for the off-resonant excitation with a frequency $\hbar \omega_L=0.35$eV far below the lowest exciton state with energy $\varepsilon = 2.03$ eV. The largest excitation field strength  $E=0.175$ V/\AA \ \ in Fig.~\ref{fig:intensity} corresponds to the driving amplitude of $f = 0.5\AA^{-1}$ or laser intensity $I=407$~GW/cm$^2$.

Figure~\ref{fig:intensity}a reveals clearly the second ($n=2$) and third ($n=3$) harmonics below the lowest excitonic resonance of $n \simeq 6$. The SHG signal is susceptible to the polarization angle, completely vanishing at $\phi = 0$.  At higher  $\omega$, the spectrum is practically independent of the angle. When the driving amplitude is small, the dependence of the SHG intensity  on the driving amplitude, shown in Fig. \ref{fig:intensity}b, follows a standard result  $I_2\propto f^{4}$ of the perturbation theory, which breaks down when $f\gtrsim 0.2\AA^{-1}$.

 The SHG intensity dependence on the driving frequency in Fig.~\ref{fig:intensity}c reveals sharp  peaks when $2 \hbar\omega_L$  coincides with exciton energy (resonance). There are different types of excitonic states that are classified similarly to atomic orbitals ($s, p \dots$)  with respect to the electron-hole coordinate difference,   with the extra complexity introduced by the valley and spin degrees of freedom. The amplitude of the resonance peaks is very sensitive to the loss rate $\gamma$ which is clearly seen when comparing the peak amplitudes with and without the loss. Away from resonances, the signal is insensitive to the loss rate.

 Further characteristics of the nonlinear response are demonstrated in Figs.~\ref{fig:intensity}d  and \ref{fig:intensity}e,  which show color density plots of the spectral intensity $I_{\omega_L,\omega}$ depending on both $\omega_L$ and $\omega$. The calculations are done for $f=0.1\AA^{-1}$  and two polarization angles $\phi=\pi/2$ and $0$. The results confirm  that the second harmonic is absent at $\phi =0$ for the off-resonant  excitation [cf. Fig. \ref{fig:intensity}a]. However, one sees that the SHG becomes visible even at $\phi =0$ when the driving frequency reaches the first excitonic resonance of $2\hbar \omega_L \simeq 2$eV. This indicates a non-trivial dependence on the polarization angle that we now explore in detail.

\begin{figure*}
    \centering
    \includegraphics[width=0.9\textwidth]{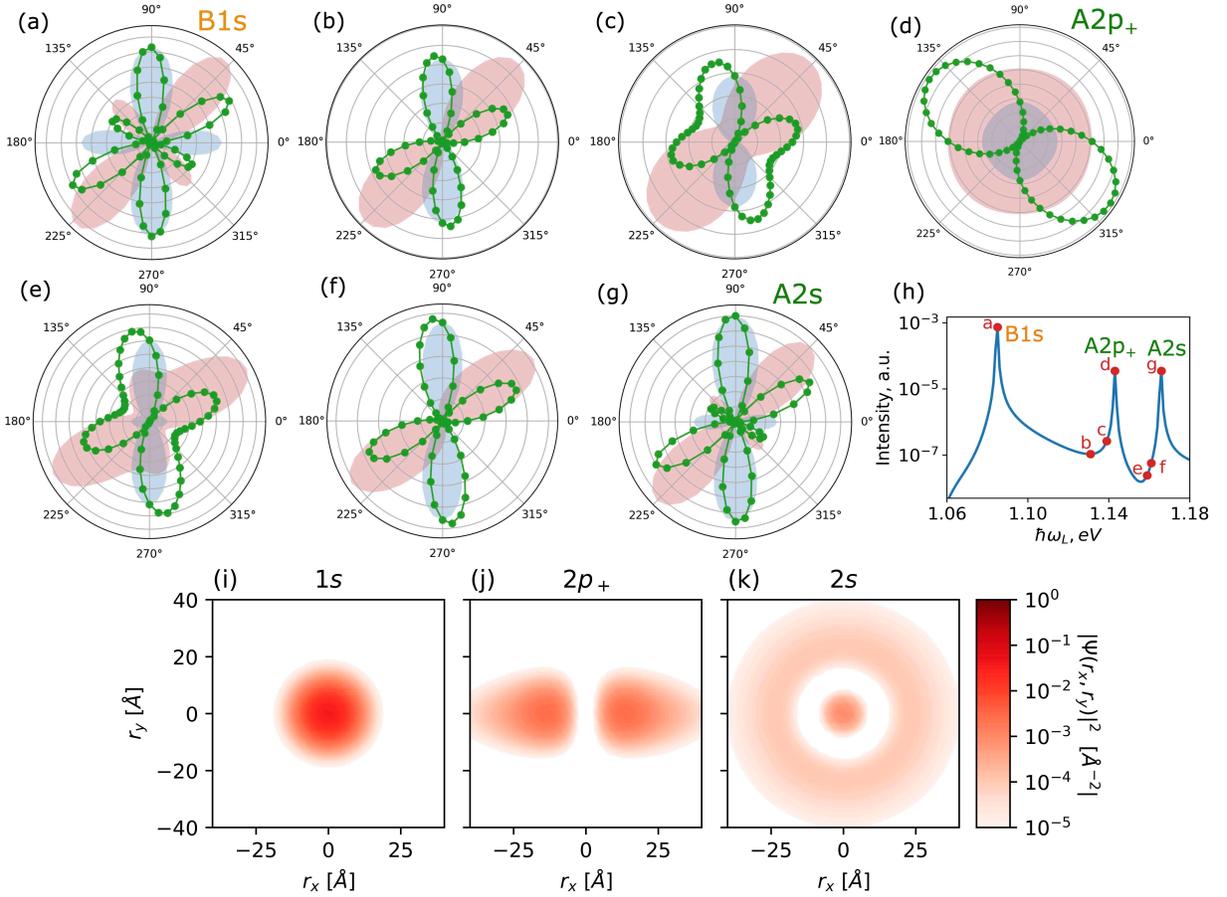}
    \caption{Angular dependence of $I$ and $|P_{x,y}|^2$ for the SHG in $\text{MoS}_2$ monolayer (a)-(g), calculated for driving field energies $\hbar\omega_L$ marked by the red points in the panel (h) [cf. Fig. \ref{fig:intensity}c]. The driving amplitude is $f=0.1\rm\;\AA^{-1}$, which corresponds to laser intensity of $I \simeq 150$ GW/cm$^2$. Panels (i-k) show spatial profiles of the exciton wave amplitude $|\Psi|^2$ as a function of the electron-hole coordinate difference, calculated for excitonic states $1s$, $2p_+$, and $2s$. 
    }
    \label{fig:SHG_polarization_2}
\end{figure*}

\subsection{Polarization diagrams}

The SHG dependence on the polarization angle is illustrated in Fig.~\ref{fig:SHG_polarization}, which plots the angular dependence of $I_2(\phi)$: here, the value of $I_2$ is given by the radial distance from the diagram center. The resulting polarization diagrams, shown in Fig.~\ref{fig:SHG_polarization},  are calculated for a standalone $\text{MoS}_2$ monolayer for selected values of $f$ and three frequencies $\omega_L$, chosen to represent off-resonant excitation, the excitation at resonance with $A1s$ and $A2p_+$ states [see Fig. \ref{fig:intensity}c]. The SHG intensity $I_2$ defined by Eq.~(\ref{eq:intensity_polarization}) depends on polarization components, $|\bar P^{x}|$ and $|\bar P^{y}|$, that are also shown in Fig.~\ref{fig:SHG_polarization} by the color shading. It is clearly seen that polarization diagrams are very sensitive  to both the amplitude $f$ and frequency $\omega_L$ of the driving field.

In the linear response limit of $f \to 0$, the polarization diagram is circular in all cases (not shown). At the same time, at the very strong driving, it reveals a familiar symmetric six-leaf pattern. This pattern is defined by the single-particle dynamics that follows the crystal symmetry of $\text{MoS}_2$, and can be used to establish the orientation of the crystal lattice in experiments ~\cite{Kumar2013,Malard2013,Li2013,Yin2014,Clark2014,Hsu2014}. Exciton states start to play a much more significant role in the dynamics for the weaker driving, which is accompanied by drastic changes in the polarization diagram.

When the excitation is off-resonant and $f$ increases, the polarization diagram first develops a  triangular shape, symmetric with respect to $120^\circ$ rotations. With the further increase in $f$, the angular dependence becomes, consecutively, first a three-leaf and then a six-leaf pattern. At still larger $f$, the six-leaf pattern becomes symmetric. Changes in the angular dependence of $I_2$ are accompanied by those in components $P^x$ and $P^y$, also shown in Fig.~\ref{fig:SHG_polarization}.

\begin{figure*}
    \centering
    \includegraphics[width=0.9\textwidth]{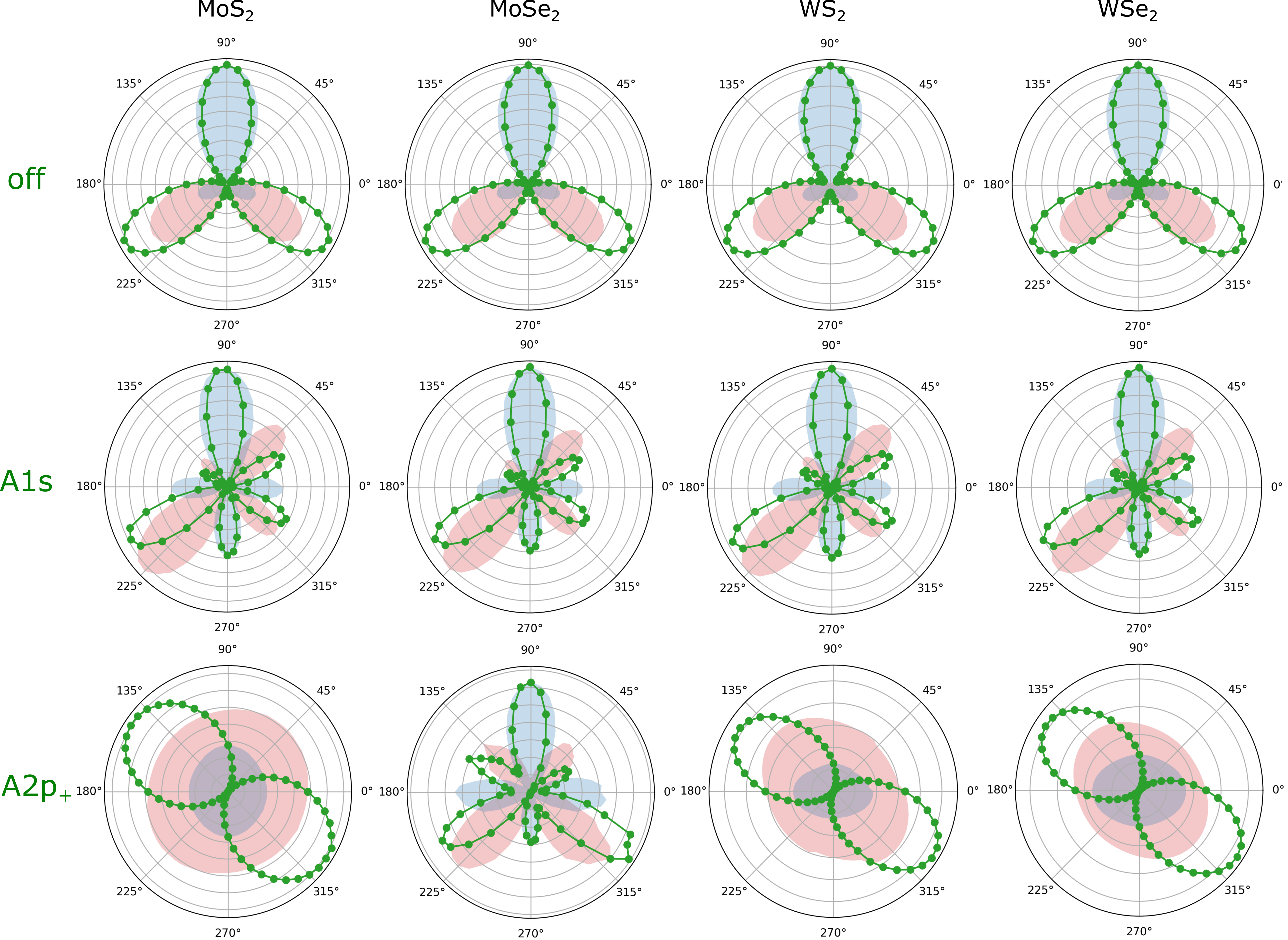}
    \caption{Angular dependence of the SHG intensity (green points) and polarization components $|P_x|^2$ (red shading) and $|P_y|^2$ (blue shading) for the driving amplitude $f = 0.1\AA^{-1}$, calculated for monolayers $\text{MoS}_2$, $\text{MoSe}_2$, $\text{WS}_2$, and $\text{WSe}_2$ (panel columns) for the off-resonant driving  with $\hbar \omega_L = 0.35$eV (top row),  at resonance with the $A1s$ state (middle row), and  at resonance with the $A2p_+$ state (bottom row).
    }
    \label{fig:SHG_polarization_materials}
\end{figure*}

When the driving frequency is at resonance with an exciton state, the polarization diagram changes qualitatively depending on the exciton type. In the $A1s$ exciton resonance, the diagram develops an asymmetric six-leaf pattern already at very small amplitudes. The same six-leaf pattern remains for all values of $f$ becoming more symmetric at stronger driving.

In contrast, in the $A2p_+$ exciton resonance in Fig.~\ref{fig:SHG_polarization}, a six-leaf pattern is seen only at $f=0.25\AA^{-1}$, while at weaker driving, the diagram differs from both the off-resonant and the $A1s$ resonance cases. For a weaker driving with $f=10^{-3}\AA^{-1}$, the angular dependence has an asymmetric two-leaf shape.  It becomes fully symmetric at larger amplitude $f=10^{-2}\AA^{-1}$. When $f$ increases, the butterfly-like pattern is formed with two equal side wings. With a further decrease in $f$, one observes four- and, then, six-leaf shapes.

Exciton-dependent differences in the polarization diagrams are  further explored by tracing how the angular dependence changes with frequency $\omega_L$. Figure \ref{fig:SHG_polarization_2} illustrates the changes by showing a diagram sequence calculated for values of $\omega_L$ in the interval between $B1s$ and $A2s$ states (the calculations are done at $f=0.1\AA^{-1}$). When $\omega_L$ increases, a non-symmetric six-leaf shape, observed at the $B1s$ resonance,  first changes into a four-leaf pattern, and then develops a butterfly-like shape with two wings, before transforming itself into a two-leaf pattern at the $A2p_+$ resonance.  The $A2p_-$ resonance does not lead to resonant enhancement of the SHG signal, because of the absence of the dipole transition to the ground state~\cite{Gong2017}. With a further increase in $\omega_L$, these transformations take place in the reversed order, producing the original asymmetric six-leaf pattern when the $A2s$ resonance is reached.

One notes that the polarization diagrams observed at  $A1s$, $B1s$ and $A2s$ resonances, shown in Figs. \ref{fig:SHG_polarization} and \ref{fig:SHG_polarization_2}, have only marginal quantitative differences. This indicates that the SHG angular dependence is similar for excitonic states of the same spatial configurations. The angular dependence at different resonances is distinguishable only when excitons have different configuration.  Spatial profiles of the exciton wave amplitude $|\Psi|^2$ for states $1s$, $2p_+$, and $2s$, as a function of the electron-hole coordinate difference ${\bf r}_e -{\bf r}_h$, are shown in Figs.~\ref{fig:SHG_polarization_2}i, j, and k, respectively. Figures  \ref{fig:SHG_polarization}  and \ref{fig:SHG_polarization_2} give a notable example of this  dependence of the state symmetry: the angular diagram observed at the $A2p_+$ resonance  deviates strongly from those at the $A1s$, $B1s$, and $A2s$ resonances. The same applies to the polarization components $P^{x,y}$ [see Fig. \ref{fig:SHG_polarization_2}] implying that this conclusion holds for arbitrary measurements setup for SHG polarization.

\subsection{Comparing different materials}

We now compare SHG polarization diagrams for monolayers of $\text{MoS}_2$, $\text{MoSe}_2$, $\text{WS}_2$, and $\text{WSe}_2$. These materials have a qualitatively similar crystal configuration and, therefore, a similar band structure. One also identifies the same types of excitonic states, i.e., $s, p$, although their spectral positions differ. When the same states are identified, one calculates the corresponding SHG polarization diagrams.

In Fig. \ref{fig:SHG_polarization_materials} the diagrams for these four materials calculated for the off-resonant driving, and the $A1s$ and $A2p_+$ resonances are plotted. One sees that the angular dependencies are indeed similar for all materials except for the $A2p_+$ resonance where $\text{MoSe}_2$ differs notably. The difference is explained by the fact that energies of the $A2p_+$ and $B1s$ states almost coincide in $\text{MoSe}_2$. This near degeneracy results in a mixture of signals typical for $B1s$ and $A2p_+$ resonances, which both contribute to the SHG and distort the angular dependence.

\section{Discussion}

A big variety of observed SHG polarization diagram types originate in the interplay of several factors affecting single- and two-particle states.  Crystal symmetry is one of the factors defining SHG angular dependence. It is taken into account by the tridiagonal warping in the Dirac model for single-particle states. The warping violates the rotational symmetry, allowing additional optical transitions. This is illustrated in the inset in Fig.~\ref{fig:intensity}c, which shows a schematic structure of the low energy exciton states contributing most to the nonlinear dynamics. A rotational symmetry admits the transitions illustrated in this scheme by solid arrows. Optical transitions cannot connect any three-state sequence, making SHG forbidden in a rotationally   invariant system, which can be formally shown, e.g., by solving the LvN equation perturbatively. The tridiagonal warping breaks the rotational symmetry allowing transitions between states $G$ and $2p_+$ (see Fig.~\ref{fig:intensity}). This creates four three-state sequences,  $G\rightarrow 1s\rightarrow 2p_+$ and $G\rightarrow 2s\rightarrow 2p_+$, for $A$ and $B$ states, respectively, giving rise to SHG.

\begin{figure}
    \centering
    \includegraphics[width=0.48\textwidth]{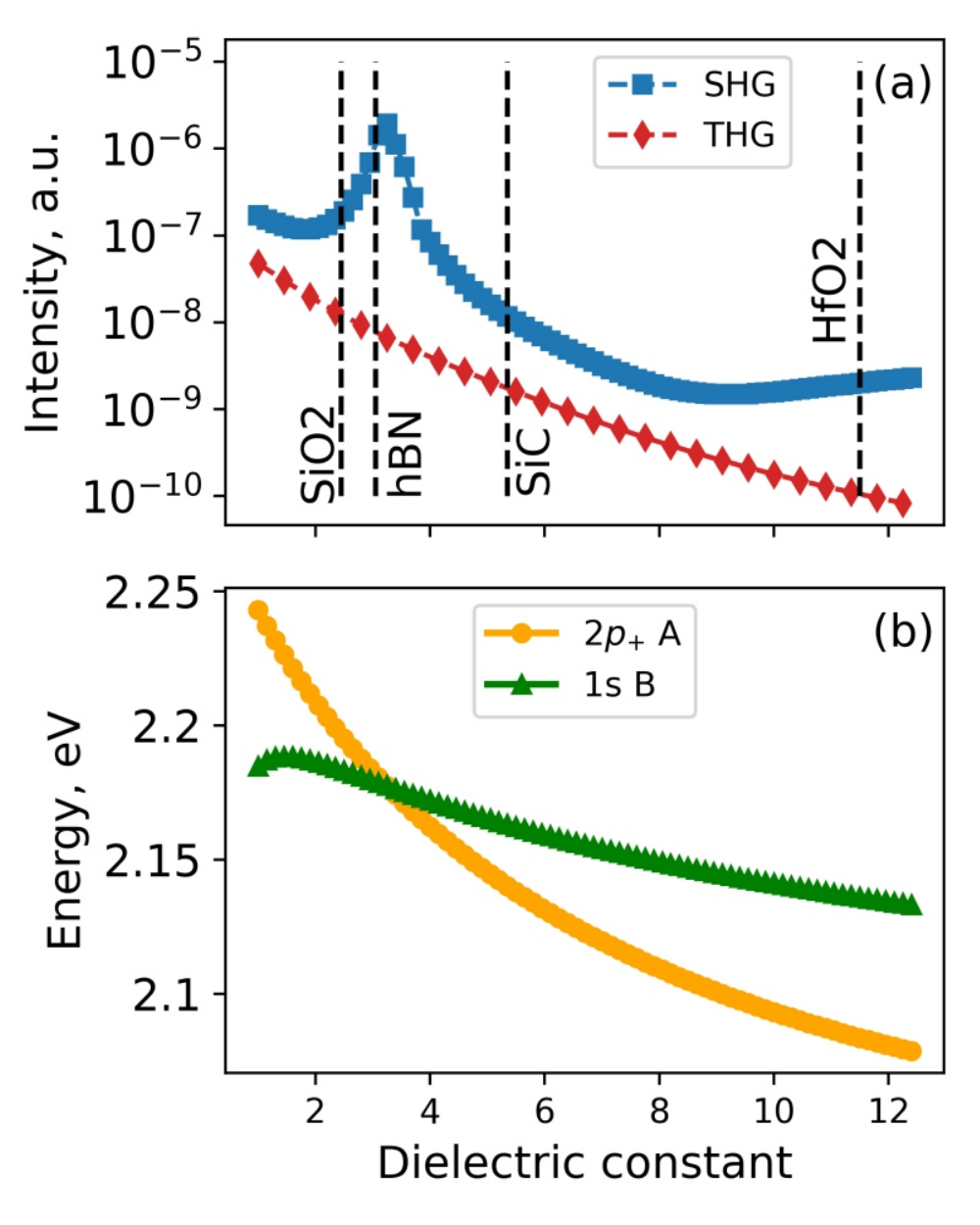}
    \caption{ (a) Intensity of the second (blue squares) and third (red diamonds) harmonics on MoS$_2$ for the driving amplitude $f = 0.1\AA^{-1}$ at resonance with the $A2p_+$ state as a function of the environment dielectric constant $\epsilon_{env}=(\epsilon_1+1)/2$, where $\epsilon_1$ is the static dielectric constant of the substrate. The dashed vertical lines correspond to $\epsilon_{env}$ values for the most commonly used substrates.  (b) Energy of $B1s$ (green triangles) and $A2p_+$ excitons (yellow circles) as a function of the environment dielectric constant.   }
    \label{fig:substrate}
\end{figure}

Another source of SHG is the quadratic term in the field-matter interaction Hamiltonian $H_f$, also introduced by the warping. The crystal symmetry is reflected in the matrix elements of $P^\alpha$ and $Q^{\alpha \beta}$ dipole operators entering $H_f$. There is a particular  relation between these matrix elements and the SHG angular dependence, which can be illustrated by estimating the polarization vector components in $H_f$. Assuming symmetric transition matrix elements, one obtains a simple expression
\begin{align}
   P_x(\phi) = \tilde f  \cos(\phi) - \alpha \tilde f^2  \sin(2\phi), \notag \\ P_y(\phi) = \tilde f  \sin(\phi) - \alpha \tilde f^2  \cos(2\phi), 
   \label{eq:polarization}
\end{align}
where $\tilde f \propto f$. By substituting this polarization vector into Eq. (\ref{eq:intensity_polarization}) and assuming that $\alpha$ is a fitting parameter, one can quantitatively reproduce all the SHG angular dependencies for the off-resonant case in Fig.~\ref{fig:SHG_polarization}. When $\tilde f$ is small, the polarization is determined by the linear contribution, such that the SHG intensity is independent of the angle. In contrast, the larger driving field activates the quadratic contribution, which dominates the angular dependency of $I(\phi)$. In this case, a symmetric six-leaf pattern emerges, as shown in Fig.~\ref{fig:SHG_polarization}. In the regime of intermediate driving amplitude, the contributions of the linear and quadratic terms are comparable, and one observes a crossover between these two extreme regimes.

However, Eq. (\ref{eq:polarization}) can be used only for the case of the off-resonant excitation.  At resonance with an excitonic state, the angular dependence becomes more complex due to a strong influence of the two-particle interactions. The configuration of an exciton state enables specific transition matrix elements, and this distorts the symmetry of the SHG angular dependence, as shown in Fig.~\ref{fig:SHG_polarization} for $A1s$ and $A2p_+$ resonances. In addition, the Coulomb interaction enhances the dipole transitions and, hence, the linear term in Eq.~(\ref{eq:Hf_exc}).

Our results show that changes induced by the exciton-related effects are most notable at the $2p_+$-state resonances, where the linear term in Eq.~(\ref{eq:Hf_exc}) is dominant in a large interval of $f$ values. The crossover between the regimes of mostly linear and mostly quadratic contributions in the light-matter interaction $H_f$ takes place at $f_c \simeq 0.2\AA^{-1}$. This coincides with the upper applicability limit for the perturbation theory result  $I\propto f^4$ in Fig.~\ref{fig:intensity}b.   The stronger the driving, the larger is the quadratic contribution, and thus the closer to the symmetric six-leaf polarization diagram.

Another important aspect to consider is the role of the exciton mixing in different valleys. A well-established theoretical fact is that Coulomb exchange coupling for excitons in different valleys is proportional to the exciton momentum~\cite{Louie2015} and vanishes in the zero-momentum limit, considered here. 
At the same time, the existence of intervalley coupling  in TMDCs is very well documented experimentally~\cite{Yu2014,Hao2016}. The strength of the intervalley coupling depends on many parameters such as doping~\cite{Nick2019}, dielectric environment~\cite{Paradisanos2020}, magnetic field~\cite{Wang2016}, and strength of disorder~\cite{Wang2013}. Besides the Coulomb exchange interaction (at finite momentum), other mechanisms such as the electron-phonon and short-range disorder interactions could lead to the intervalley coupling, and there are ongoing debates as to which of those is the dominant one~\cite{Yu2014b,Glazov2014,Schaibley2016,Glazov2017,Paradisanos2020}. Since we deal with doubly degenerate states, the interaction between them of any infinitesimal strength would mix the wavefunctions of those states by 50\%, even though the energies of those states would not change much. We performed controlled calculations, where doubly degenerate states originating from different valleys are mixed by 50\%, i.e., $\Psi_{new1}=(\Psi_{old1}+\Psi_{old2} )/ \sqrt{2}$, $\Psi_{new2}=(\Psi_{old1}-\Psi_{old2} )/ \sqrt{2}$, where old and new subscripts refer to the old and new wavefunctions, correspondingly.  When the matrix elements entering the Hamiltonian for the Liouville - von Neumann equation use new wavefunctions, we do not find changes in the resulting angular polarization diagrams reported earlier. Therefore, our main conclusions on the angular polarization diagram dependence on the excitation power and resonant conditions are not sensitive to the $K-K’$ valley exciton mixing.  

Finally, we discuss the influence of the dielectric environment on the excitonic effects in the nonlinear response. The environment affects the effective dielectric constant of the system, modifying the strength of the Coulomb interaction, the  binding energy of excitons, and, thus, their contribution to the dynamics. This is illustrated in Fig.~\ref{fig:substrate}a, which plots the intensity of the second and third harmonics at resonance with the $A2p_+$ state as functions of the environment dielectric constant $\epsilon_{env}$. As $\epsilon_{env}$ increases from 1 to 10, the intensity of both harmonics decreases by more than two orders of magnitude.  A sharp peak in the SHG signal at $\epsilon_{env} \simeq 4$ in Fig.~\ref{fig:substrate}a appears due to the degeneracy of the $A2p_+$ and $B1s$ energies at this point [see Fig.~\ref{fig:substrate}b].

In order to observe the polarization pattern change, the field amplitude has to be reduced by an order of magnitude under off-resonant conditions (see the top panels in Fig.~\ref{fig:SHG_polarization}), which in turn translates into a four-orders-of-magnitude reduction in the SHG signal (see Fig.~\ref{fig:intensity}b).  This creates a tremendous challenge in the experiment to observe an ultra-low SHG signal. A much more favorable situation occurs when the excitation laser energy resonates  with half the energy of the $A2p_+$ exciton (see the bottom panels in Fig.~\ref{fig:SHG_polarization}). In this case, a reduction of a factor of two in field amplitude or only one order of magnitude in laser intensity is needed to observe substantial  changes in the polarization pattern. Since the exciton energies are sensitive to doping and the dielectric environment, the most favorable measurement setup would involve scanning the excitation laser wavelength at fixed laser power. According to Fig.~\ref{fig:SHG_polarization_2}, a six-leaf pattern would evolve into a four-leaf pattern and a two-leaf pattern under a realistic laser intensity of about 150 GW/cm$^2$. Thermal management can be improved by choosing dielectric substrates with high thermal conductivity, such as diamond, to avoid sample burning.  However, suspended samples can enable measurements in the transmission mode with the use of two different polarizers for the excitation laser energy and its second harmonic overturn.   According to Fig.~\ref{fig:substrate}a, a reduced dielectric screening of the environment gives a two-orders-of-magnitude enhancement of the SHG signal, which significantly reduces the length of time of measurements that is needed for a good signal-to-noise ratio.

\section{Conclusions}

Our work demonstrates that SHG in TMDC monolayers is notably affected by the many-particle exciton effects. The SHG signal increases by orders of magnitude when the driving pulse is at resonance with an excitonic state. More importantly, excitons  alter the SHG polarization angular dependence  qualitatively. Its dependence on the energy and amplitude of the driving field undergoes significant changes depending on the type of the resonating exciton. The influence of excitons gives rise to deviations from the symmetric six-leaf angular dependence in monolayers with an undistorted crystal structure.  This conclusion is generic, being supported by the qualitatively similar polarization patterns and their changes obtained for $\text{MoS}_2$, $\text{MoSe}_2$, $\text{WS}_2$, and $\text{WSe}_2$. 

Our results open a pathway to a nonlinear spectroscopy method to probe exciton states by the symmetry of the polarization angular dependence of the SHG signal measured as a function of the excitation strength. Indeed, nonlinear spectroscopy can probe Rydberg states of strongly bound excitons in a one-dimensional  system~\cite{FengWang2005,Maultzsch2005}, not accessible in linear optical absorption. Two-dimensional materials offer a rich diversity of optical signatures in photoluminescence spectra~\cite{Wang2018}; the origin of some of them is still not identified.  The SHG polarization angular dependence as a function of excitation power offers a powerful characterization tool to unravel the nature of strongly correlated excited states in low-dimensional materials in addition to the commonly used magnetic-field- and electric-field-dependent optical spectroscopy. 

\vspace{1 cm}

\section*{Acknowledgments}

We gratefully acknowledge Tony Heinz (Stanford University) for drawing our attention to the nonlinear properties of TMDCs, Vasily Kravtsov (ITMO University) and John Schaibley (University of Arizona) for informative and insightful discussions of the experimental challenges in the realization of the proposed type of spectroscopy.
Calculation of the exciton states used in the nonlinear model was supported by the Russian Science Foundation under Grant No. 18-12-00429. The study of the influence of the dielectric environment 
was supported by MEPhI Program Priority 2030 
and performed with the help of the
NRNU MEPhI high-performance computing center. Y.\,V.\,Z. is grateful to Deutsche Forschungsgemeinschaft (DFG, German Research Foundation) SPP 2244 (Project No. 443416183) for the financial support.
V.\,P.  acknowledges computational facilities at the Center for Computational Research at the University at Buffalo (\url{http://hdl.handle.net/10477/79221}).

\bibliography{bibliography}

\end{document}